\def\mc{meridional circulation}
\def\wald{Waldmeier effect}
\def\bl{Babcock--Leighton}

\documentclass{iaus}
\usepackage{graphicx}

\title[Modeling irregular solar cycle] 
{Is meridional circulation important in modelling irregularities of the solar cycle?}

\author[Karak \& Choudhuri]   
{Bidya Binay Karak
 \and Arnab Rai Choudhuri}

\affiliation{Department of Physics, Indian Institute of Science, Bangalore 560012, India \\ email: {\tt bidya$\_$karak@physics.iisc.ernet.in}\\[\affilskip]}

\pubyear{2011}
\volume{286}  
\jname{Comparative Magnetic Minima: characterizing quiet times in the Sun and stars}
\editors{A.C. Editor, B.D. Editor \& C.E. Editor, eds.}
\begin{document}

\maketitle

\begin{abstract}
We explore the importance of meridional circulation variations in modelling 
the irregularities of the solar cycle by using the flux transport dynamo model.
We show that a fluctuating meridional circulation 
can reproduce some features of the solar cycle  
like the Waldmeier effect 
and the grand minimum. However, we get all these results only 
if the value of the turbulent diffusivity in the convection zone 
is reasonably high.
\keywords{Sun: dynamo --- Sun: activity --- sunspots}
\end{abstract}

\firstsection 
\section{Introduction}
The solar cycle is not regular. There were several grand minima like the Maunder minimum
during which the activity level was extremely low.
Although the solar activity has varied approximately cyclically since the Maunder minimum, the 
amplitudes and the periods of individual cycles varied in an irregular manner.
Another important feature of solar cycles is the Waldmeier effect, which is basically the
anti-correlation between the rise time and the peak sunspot number.
However, we define two different aspects of 
it, which we call WE1 and WE2. By WE1 we refer to the anti-correlation between 
the \textit{rise time} and the peak sunspot number, whereas by WE2 we refer to the positive 
correlation between the \textit{rise rate} and the peak sunspot number.

Our motivation is to model the irregularities of the solar cycle, including features like
the Waldmeier effect, by using the flux transport dynamo model (see Choudhuri 2011 and references therein). 
In this model, the toroidal field is 
generated near the base of the convection zone by differential rotation and 
the poloidal field is generated near the solar surface by the decay
of tilted bipolar sunspots. These two source regions are connected to
each other by several transport agents. One important flux transport agent
in this model is the turbulent magnetic diffusivity ($\eta_t$). However, its
value in the whole convection zone is not properly constrained. This has led
to two different classes of models, in which the diffusivity has been taken
to be high or low (see Jiang, Chatterjee \& Choudhuri 2007; Choudhuri 2011). 
The values of $\eta_t$
in these two classes of models are taken in the ranges $\sim 10^{12}-10^{13}$ 
cm$^2$ s$^{-1}$ and $\sim 10^{10}-10^{11}$ cm$^2$ s$^{-1}$ respectively.
Another important flux transport agent in this model is the
meridional circulation. The present understanding of its origin --- and, more importantly,
its fluctuations --- is very primitive. It is believed that the 
the meridional circulation arises from a slight imbalance between two large terms 
(the centrifugal force due to the variation of angular velocity with 
distance from the equatorial plane and the thermal wind due to a temperature variation with latitude).
Therefore, we expect that there may be random variations in the \mc\
due to stochastic fluctuations in any one of these driving forces.
Only since 1990s we have some observational data of \mc\ near the surface. 
Therefore we do not know whether it had large variations in the past. However, 
we can get some idea of 
the variations of the \mc\ from the observed periods of the past solar cycles.
In the next section, we discuss how this is done by using the flux transport dynamo
model and then we model solar cycles using the variable meridional circulation.

\section{METHODOLOGY AND RESULTS}

\begin{figure}
\begin{center}
\includegraphics[width=1.00\textwidth]{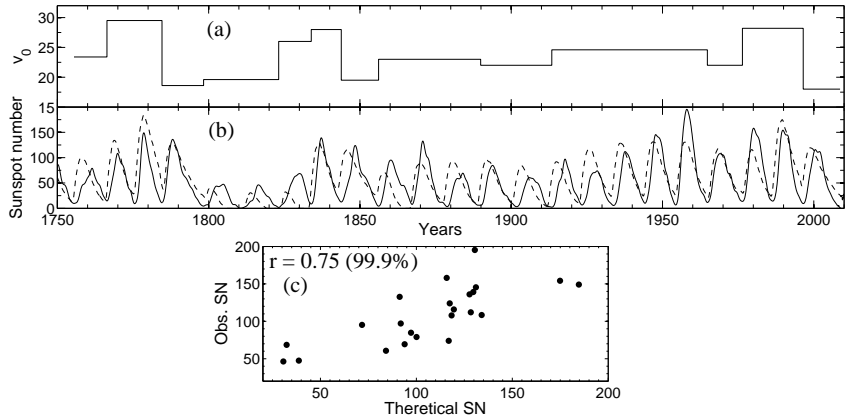}
\caption{(a) Variation of amplitudes of \mc\ $v_0$ (in m~s$^{-1}$) with time 
(in yr). The solid line is the 
variation of $v_0$ used to match the theoretical periods with the 
observed periods. (b) Variation of theoretical sunspot number (dashed 
line) and observed sunspot number (solid line) with time. (c) Scatter diagram showing 
peak theoretical sunspot number and peak observed 
sunspot number. From Karak (2010).}
\label{fit23}
\end{center}
\end{figure}

We know that the period of the cycle in the flux transport dynamo model
is primarily determined by the strength of 
the \mc\  (Dikpati \& Charbonneau 1999). A stronger \mc\ makes
the dynamo period shorter.
Therefore it should be possible to match the observed periods
of the last 23 solar cycles by using a variable \mc. Karak (2010) performed this experiment 
using a high diffusivity model based on the model of Chatterjee, Nandy \& Choudhuri (2004).
Fig.~\ref{fit23}(a) shows the variation of the amplitude of \mc\ required to match
the periods of last 23 solar cycles. From this figure, we see that the \mc\ varied 
significantly with time. Therefore, if the flux transport dynamo model is the correct 
model for the solar cycle, then we have to conclude that the \mc\ had large variations
in the past.
Now let us look at the variation of theoretical sunspot number obtained by Karak (2010)  shown by the
dashed line in Fig.~\ref{fit23}(b). For comparison, the observed sunspot number is shown 
by the solid line.
Surprisingly, most of the theoretical sunspot cycle amplitudes
are matched with the observed ones. We do get a good correlation between these
two as shown in Fig.~\ref{fit23}(c). This is a very important result of this
analysis because our motivation was only to match the solar cycle periods and to get
some idea of the variation of \mc\ in the past. However, while doing this, we find 
that most of the solar 
cycle amplitudes are also matched to some extent. Therefore, this study
suggests that a significant amount of fluctuations in the strengths of the
cycles is arising from the variations in the \mc, which seems 
important in modelling not only the solar cycle periods but also the 
cycle amplitudes.

Now let us explain the physics of this result based on the arguments given by
Yeates, Nandy \& Mackay (2008).
We know that in the flux transport dynamo, the
production of the toroidal field is more if the poloidal field remains
in the tachocline for longer time and vice versa. However, the poloidal field also
diffuses during its transport through the convection zone. As a result, if
the diffusivity is very high, then much of the poloidal field diffuses away
and not much of it reaches the tachocline to induct the toroidal field.
Therefore, when we decrease $v_0$ in high diffusivity model to match
the period of a longer cycle, the poloidal field gets more time to diffuse
during its transport through the convection zone. This ultimately leads
to a lesser generation of toroidal field and hence the cycle becomes
weaker. On the other hand, when we increase the value of $v_0$ to match 
the period of a shorter cycle, the poloidal field does not get much 
time to diffuse in the convection zone. Hence it produces stronger 
toroidal field and the cycle becomes stronger.
Consequently, we get weaker
amplitudes for longer periods and vice versa. However, this is not the case 
in low diffusivity model because in this model the diffusive decay of the fields is not
very important. As a result, the slower $v_0$ means that
the poloidal field remains in the tachocline for longer time and therefore
it produces more toroidal field, giving rise to a strong cycle. Therefore, we do not get
the correct correlation between the amplitudes of theoretical sunspot number and
those of observed sunspot number when we repeat the same analysis in the low diffusivity 
model.

\begin{figure}[!h]
\begin{center}
\includegraphics[width=1.00\textwidth]{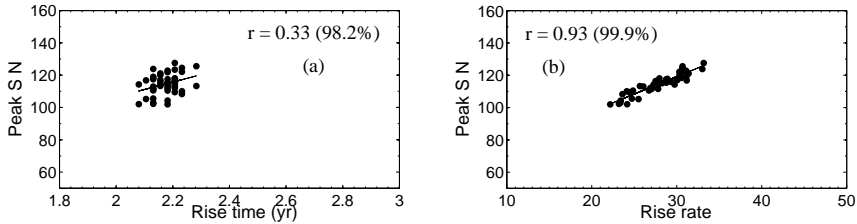}
\caption{Results for WE1 (left panel) and WE2 (right panel) obtained
by introducing fluctuations in the poloidal field at the minima. From
Karak \& Choudhuri (2011).}
\label{pol}
\end{center}
\end{figure}

\begin{figure}[!h]
\begin{center}
\includegraphics[width=1.00\textwidth]{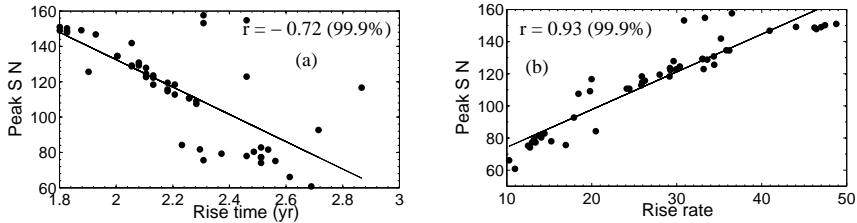}
\caption{Same as earlier but obtained by introducing fluctuations
in the meridional circulation. From Karak \& Choudhuri (2011).}
\label{mc}
\end{center}
\end{figure}
Next we present some results from Karak \& Choudhuri (2011), who studied 
the \wald\ using the flux transport dynamo model.  
The stochastic fluctuations in the \bl\ process of generating poloidal field and the stochastic fluctuations 
in the \mc\ are the two main sources of irregularities in this model. 
Therefore, to study the \wald, we first introduce suitable stochastic fluctuations 
in the poloidal field source term. Fig.~\ref{pol} shows the result.
We see that this study cannot reproduce WE1 (Fig.~\ref{pol}(a)), although it 
reproduces WE2 (Fig.~\ref{pol}(b)). Next we introduce stochastic fluctuations in 
the meridional circulation. Fig.~\ref{mc} shows this result. We 
see that both WE1 and WE2 are remarkably reproduced in this case. However, 
when we repeat the same study with the low diffusivity model of Dikpati \& 
Charbonneau (1999), we fail to reproduce WE1. Only WE2 is 
reproduced in this model.

\begin{figure}[!h]
\begin{center}
\includegraphics[width=1.0\textwidth]{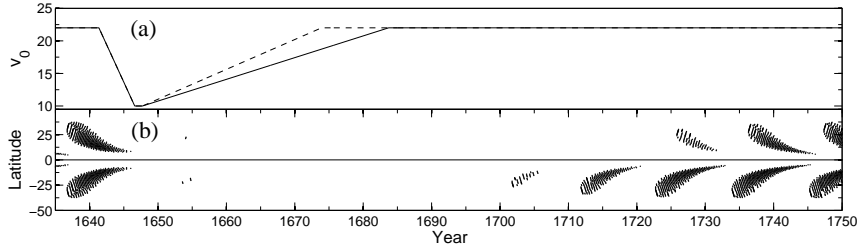}
\vspace{-5.8cm}
\caption{(a) The solid and dashed line
show the variations of $v_0$ (in m~s$^{-1}$) in northern and southern hemispheres with time. (b) The butterfly diagram.
From Karak (2010).}
\label{mm}
\end{center}
\end{figure}

Karak (2010) found that a decrease of the \mc\ to a very low
value can reproduce a Maunder-like grand minimum.
Fig.~\ref{mm}(a) shows the variation of $v_0$ required to
model a Maunder-like grand minimum. 
Fig.~\ref{mm}(b) shows the butterfly diagram of sunspot numbers.
This result reproduces most of the features of the 
Maunder minimum remarkably well.

We have shown that the temporal variation of the meridional circulation is important in modelling 
irregularities of the solar cycle. Our studies also suggest that
there have been large variations in the \mc\ in the past.
With suitable stochastic fluctuations in the meridional circulation, 
we are able to reproduce many important 
irregular features of the solar cycle including the Waldmeier effect and 
Maunder-like grand minima. However we fail to reproduce these
results in a low diffusivity model. 
Therefore this study along with some other 
studies (Chatterjee, Nandy \& Choudhuri 2004; Chatterjee \& Choudhuri 2006; 
Jiang, Chatterjee \& Choudhuri 2007; Choudhuri \& Karak 2009; Goel \& Choudhuri 2009; Hotta \& Yokoyama 2010;
Karak \& Choudhuri 2012) 
supports the high diffusivity model for the solar cycle.

\section*{Acknowledgments}
We thank IAU funding agencies and DST, India for financial support.

\end{document}